\documentclass[11pt]{article}
\usepackage{epsfig,amsmath,amssymb,amsbsy}
\usepackage{amsfonts}
\usepackage{times}

\renewcommand{\Re}{{\text{Re\,}}}
\renewcommand{\Im}{{\text{Im\,}}}

\newcommand{\RR}{\mathbb{R}}

\newcommand{\pair}{(\kappa,\omega)}
\newcommand{\pairzero}{(\kappa_0,\omega_0)}
\newcommand{\om}{\varpi}
\newcommand{\ka}{{\tilde\kappa}}
\renewcommand{\O}{{\cal O}}
\newcommand{\Oko}{{\cal O}(|\ka|+|\om|)}

\newcommand{\ppsi}{{\boldsymbol{\psi}}}
\newcommand{\pphi}{{\boldsymbol{\phi}}}

\newcounter{entry}

\begin{document}

\bibliographystyle{plain}

\title{Fano-Type Resonance of Waves in Periodic Slabs}

\author{ Natalia Ptitsyna \\ Louisiana State University
\and
Stephen P. Shipman \\ Louisiana State University
\and
Stephanos Venakides \\ Duke University
}
\maketitle

\begin{abstract}
We investigate Fano-type anomalous transmission of energy of plane waves across lossless slab scatterers with periodic structure in the presence of non-robust guided modes.  Our approach is based on rigorous analytic perturbation of the scattering problem near a guided~mode and applies to very general structures, continuous and discrete.
\end{abstract}

\section{Introduction}

Resonances with sharp peaks and dips have been observed in the graphs of the frequency dependence of the transmission of energy of plane waves through periodic slab structures. \cite{EbbesenLezecGhaemi1998,Ptitsyna2008,ShipmanVenakides2003,TikhodeevYablonskiMuljarov2002}. These anomalies have been connected to {\em isolated real points} on the complex dispersion relation $D(\kappa,\omega)\!=\!0$ that relates the frequency $\omega$ of a sourceless field to its Bloch wave vector $\kappa$ along the slab \cite{Ptitsyna2008,ShipmanVenakides2003}.  The real point of the dispersion relation corresponds to a true guided mode in the slab, that is, one that decays exponentially away from the slab.  Because the point is isolated, the mode is {\em non-robust} with respect to perturbations of $\kappa$, $\omega$, or the physical or geometric parameters of the structure.  This feature gives rise to sharp transmission anomalies near the frequency of the guided mode, as shown in Figs. 1 and 2.

The anomalies observed resemble those investigated by U. Fano \cite{Fano1961} in the context of quantum mechanics, in which an embedded eigenvalue is ``dissolved" into the continuous spectrum upon perturbation of the system, resulting in resonant dynamical behavior.  The shape that he derived for the resonance involves two parameters and turns out to be a special case of a more general shape derived rigorously for the graph of energy transmission through photonic crystal slabs with a symmetry assumption on the slab geometry \cite{ShipmanVenakides2005}.
Typically, symmetry gives rise to standing guided waves ($\kappa\!=\!0$), as discussed in \cite{Bonnet-BeStarling1994,ShipmanVolkov2007}, and only standing waves were analyzed in \cite{ShipmanVenakides2005}.  In the present work, we compute anomalies near traveling guided modes ($\kappa\!\not=\!0$).
The frequency of a non-robust guided mode is an embedded eigenvalue for a fixed value of $\kappa$.
The dissolution of the eigenvalue into the continuous spectrum corresponds to the destruction of the guided mode when $\kappa$ is perturbed.
We demonstrate the results through two examples, the scattering of polarized EM fields by a lossless dielectric structure and the scattering of waves in a two-dimensional lattice by a one-dimensional periodic lattice attached along a line.

\section{Analytic connection of scattering states and guided modes}

In \cite{ShipmanVenakides2005}, we reformulate the Maxwell equations in a photonic crystal slab (Fig. 2) at fixed frequency $\omega$ and Bloch wavenumber $\kappa$ (the latter necessarily parallel to the slab) as a  boundary-integral equation (see \cite{Muller1969,Nedelec2000}, for example). The equation
\begin{equation}\label{inhomogeneous}
  A(\kappa,\omega)\ppsi = \pphi
\end{equation}
relates two fields $\pphi$ and $\ppsi$ defined on the interface between the scatterer material (taken to be homogeneous and isotropic) and the ambient medium.
The field $\ppsi$ on the interface  contains information that is necessary and
sufficient for the calculation of the EM field in the whole space; the calculation consists of the evaluation of an integral.  Similarly, the field $\pphi$ contains information for the calculation of the incident field. Solving the above equation for $\ppsi$ is equivalent to calculating the field in the photonic crystal and in the ambient medium established as a result of the source field $\pphi$.  We call $\ppsi$
the total field to distinguish it from the scattered field, which is the difference between
the total and incident fields in the ambient medium; clearly, a scattered field is required to satisfy a radiation condition. The integral operator $A$ is of second-kind Fredholm type with index zero, thus, it either has a bounded inverse or it has a nullspace of finite dimension and a range of the same codimension.

The integral operator $A$ is parametrized by $\omega$ and $\kappa$. When
either of these takes a nonreal value, a field $\ppsi$, bounded on the interface, may
produce an unbounded and hence unphysical field in space due to  growth of the exponential $e^{i\kappa x}$ or $e^{i\omega t}$.  Such fields are not only mathematically but also physically useful and form the foundation of the so-called leaky modes. A (generalized) guided mode is a nonzero solution of the sourceless equation
\begin{equation}\label{homogeneous}
  A(\kappa,\omega)\ppsi = \boldsymbol{0}.
\end{equation}
The pairs $\pair$ for which such a solution exists satisfy a dispersion relation $D\pair = 0$.  For pairs $\pair$ such that $\kappa$ is real and $D\pair=0$, we must have $\Im\omega\leq0$ \cite{Ptitsyna2008,ShipmanVenakides2003}.  In the case that $\Im\omega=0$, a solution to \eqref{homogeneous} represents an EM field that decays exponentially with distance from the scatterer and therefore represents a true guided mode.
We shall work with a simple eigenvalue branch $\ell\pair$ of $A\pair$, for which we necessarily have that $\ell\pair=0$ implies $D\pair=0$.

The framework we have described explicitly for the Maxwell equations arises for very general dielectric and metal structures (not necessarily with homogeneous components) and for other continuous and discrete problems of scattering by a slab; the methods of arriving at the formulation vary from problem to problem.   For waves in an $n$-dimensional uniform lattice scattered by a $(n\!-\!1)$-dimensional periodic lattice (Fig. 1), the problem is reduced to a finite-dimensional one, and $A$ is a matrix.  

%
%

For the purpose of investigating anomalous scattering behavior near a guided mode, we let $\pairzero$ be a real pair that satisfies $\ell\pairzero=0$ and use as an incident source field the plane wave
\begin{equation}
  \phi_\text{inc}(x,z) = \ell\pair e^{i\kappa x}e^{i\eta z} \qquad \text{(incident from the left),}
\end{equation}
where $x$ is the multi-variable in the directions parallel to the slab and $z$ is the single variable perpendicular to the slab ($x$ and $z$ may be discrete or continuous).  The number $\eta$ depends on $\kappa$, $\omega$, and the structure itself.  We shall work in the regime of only one propagating diffractive order, that is, one Fourier harmonic in the periodic variable $x$ that carries energy in the $z$-direction.  In this case, the reflected and transmitted fields (at far field) are simple:
\begin{eqnarray}
  &&  \phi_\text{refl}(x,z) = a\pair e^{i\kappa x}e^{-i\eta z} \qquad \text{(reflected on the left),} \\
  &&  \phi_\text{trans}(x,z) = b\pair e^{i\kappa x}e^{i\eta z} \qquad \text{(transmitted on the right).}
\end{eqnarray}
It can be shown that the coefficients $a\pair$ and $b\pair$ are analytic functions of $\pair$ near $\pairzero$ 
\cite{Ptitsyna2008,ShipmanVenakides2005}.  At pairs $\pair$ that satisfy $\ell\pair=0$, the the right-hand side of \eqref{inhomogeneous} is zero, and the solution is a guided mode.  Thus we obtain an analytic connection between scattering states and generalized guided modes near the true guided mode at $\pairzero$.

\section{Analysis of transmission near a guided mode frequency}


The analysis of the resonant transmission shape near a guided mode is based on the analysis of the three analytic functions $\ell\pair$, $a\pair$, and $b\pair$
near the real parameters $\pairzero$ of a guided mode.  The mode condition $\ell\pairzero = 0$, together with that of conservation of energy, namely $|\ell|^2 = |a|^2 + |b|^2$ for real values of $\kappa$ and $\omega$, implies that $\pairzero$ is a root of the three analytic functions $\ell$, $a$, and $b$ simultaneously:
\begin{equation}
  \ell\pairzero = 0, \quad
  a\pairzero = 0, \quad
  b\pairzero = 0, \quad
  \text{at } \pairzero \in\RR^2.
\end{equation}
The perturbation analysis of $\ell$, $a$, and $b$ near this common real root relies principally on the following two conditions, which we have already mentioned:
\begin{eqnarray}
  && |\ell\pair|^2 = |a\pair|^2 + |b\pair|^2 \text{ for real $\kappa$ and $\omega$,} \label{energy}\\
  && \text{if } \ell\pair = 0 \text{ for } \kappa\in\RR, \text{ then } \Im\omega\leq0. \label{lhp}
\end{eqnarray}
The following conditions are assumed because they hold generically:
\begin{equation}
   \frac{\partial \ell}{\partial \omega}\pairzero \not= 0, \quad
   \frac{\partial a}{\partial \omega}\pairzero \not= 0, \quad
   \frac{\partial b}{\partial \omega}\pairzero \not= 0.
\end{equation}
The Weierstra\ss\ preparation theorem for analytic functions of several variables \cite{Markushevi1965} then provides the following forms (with $\om = \omega-\omega_0$ and $\ka = \kappa-\kappa_0$):
\begin{eqnarray}
  && \ell\pair = e^{i\theta_1}\left[\om + \ell_1\ka + \ell_2\ka^2 + \O(\ka^3)\right]\left[1+\Oko\right], \label{form1E}\\
  && a\pair = e^{i\theta_2}\left[\om + r_1\ka + r_2\ka^2 + \O(\ka^3)\right]\left[r_0+\Oko\right], \label{form1a}\\
  && b\pair = e^{i\theta_3}\left[\om + t_1\ka + t_2\ka^2 + \O(\ka^3)\right]\left[t_0+\Oko\right]. \label{form1b}
\end{eqnarray}
The function $\ell$ has been normalized so that the constant in the second factor is equal to~1.  All factors in these forms are analytic, and the numbers $r_0$ and $t_0$ are real and positive.

Consequences of \eqref{lhp} are that $\ell_1$ is real and $\Im\ell_2\geq0$:
\begin{equation}
  r_0>0, \quad t_0>0, \quad \ell_1\in\RR, \quad \Im\ell_2\geq0.
\end{equation}
Using the forms (\ref{form1E},\ref{form1a},\ref{form1b}), we can examine various terms in the expansion of the equation of conservation of energy \eqref{energy} for real values of $\kappa$ and $\omega$.
Computation of $|\ell|^2$ yields
\begin{multline}
  \ell\bar \ell = \left[ \om^2 + \ell_1^2 \ka^2 + 2\ell_1\om\ka + 2\Re\ell_2\om\ka^2 + 2\ell_1\Re\ell_2\ka^3 + \right. \\
  \left. + \,(2\ell_1\Re\ell_3 + |\ell_2|^2)\ka^4 + ... \right] \left[ 1+\Oko \right],
\end{multline}
and computation of $|a|^2$ and $|b|^2$ yield analogous expressions.
In the case that $\ell_1\not=0$, three terms of the expansion of $|\ell|^2 = |a|^2 + |b|^2$ are determined in terms of the coefficients expressed explicitly in the forms (\ref{form1E},\ref{form1a},\ref{form1b}).  In the case that $\ell_1=0$, some of these terms become trivial and others become relevant.  We analyze these two cases separately.

\subsection{Case 1: $\ell_1\not=0$}

The three terms in the expansion of \eqref{energy} that give information about the coefficients in the forms (\ref{form1E},\ref{form1a},\ref{form1b}) are
\begin{equation}\label{relations1}
\renewcommand{\arraystretch}{1.2}
\left.
  \begin{array}{ll}
      \om^2 \;\text{term:} & 1 = r_0^2 + t_0^2, \\
      \ka^2 \;\text{term:} & \ell_1^2 = r_0^2\,|r_1|^2 + t_0^2\,|t_1|^2, \\
      \om\ka \;\text{term:} & \ell_1 = r_0^2\,\Re r_1 + t_0^2\,\Re t_1.
  \end{array}
\right.
\end{equation}
Because of the convexity of the function $x\mapsto x^2$, the $\om^2$-term and the $\om\ka$-term imply that
\begin{equation}\label{one}
  \ell_1^2 \leq r_0^2\,(\Re r_1)^2 + t_0^2\,(\Re t_1)^2,
\end{equation}
with equality if and only if $\Re r_1 = \Re t_1$, and the $\ka^2$-term is expanded to
\begin{equation}\label{two}
  \ell_1^2 = r_0^2\,(\Re r_1)^2 + t_0^2\,(\Re t_1)^2 + r_0^2\,(\Im r_1)^2 + t_0^2\,(\Im t_1)^2.
\end{equation}
From conditions \eqref{one} and \eqref{two} together, we infer that
\begin{equation}
  \Im r_1 = \Im t_1 = 0  \quad\text{and}\quad  r_1 = t_1 = \ell_1.
\end{equation}
We thus obtain useful expressions for the zero-sets of $\ell$, $a$, and $b$ near the guided mode pair $\pairzero$:
\begin{eqnarray}
  \ell\pair=0 &\iff& \omega = \omega_0 - \ell_1(\kappa-\kappa_0) - \ell_2(\kappa-\kappa_0)^2 - \dots\,, \label{form2E}\\
  a\pair=0 &\iff& \omega = \omega_0 - \ell_1(\kappa-\kappa_0) - r_2(\kappa-\kappa_0)^2 - \dots\,, 
  \quad \text{($\ell_1\in\RR$)} \label{form2a}\\
  b\pair=0 &\iff& \omega = \omega_0 - \ell_1(\kappa-\kappa_0) - t_2(\kappa-\kappa_0)^2 - \dots\,. \label{form2b}
\end{eqnarray}
We observe that, if all coefficients $r_n$ and $t_n$ vanish, then the curve $a\pair=0$ (resp. $b\pair=0$), for real values of $\kappa$, describes real frequencies $\omega$ for which transmission $T$ is equal to $100\%$ (resp. $0\%$).  We do know that $\ell_1$ is real, so that these curves lie at most a distance $\O((\kappa-\kappa_0)^2)$ from the real $\omega$-axis, and one can deduce that the real parts of the frequencies on these curves correspond to peaks and dips of the transmission, which do not necessarily reach exactly $100\%$ or~$0\%$.

These results allow us to make two important observations about the shape of the transmission resonance as a function of frequency.  Both of these are illustrated on the graphs in Fig. 1.
\begin{enumerate}
  \item For $\kappa\not=\kappa_0$, either both the peak and dip of $T$ as a function of $\omega$ lie to the left of the frequency $\omega_0$ or both lie to the right of $\omega_0$.  On which side they lie depends on the sign of $\ell_1$ and $\kappa-\kappa_0$. 
  \item The order in which the peak and dip in $T$ occur on the real $\omega$-axis is the same for $\kappa<\kappa_0$ as it is for $\kappa>\kappa_0$ (assuming $r_2\not=t_2$).  This is because the coefficients of the linear terms in \eqref{form2a} and \eqref{form2b} are equal.  If $r_2<t_2$, then the peak comes to the right of the dip, and if $t_2<r_2$, then the peak comes to the left of the~dip.
\end{enumerate}

The transmission coefficient (square root of transmitted energy) as a function of real $\kappa$ and $\omega$~is
\begin{multline}
  T(\kappa,\omega) = \left|\frac{b\pair}{\ell\pair}\right| = 
  t_0\frac{|\om + \ell_1\ka + t_2\ka^2 + \cdots|}{|\om + \ell_1\ka + \ell_2\ka^2 + \cdots|}
  | 1 + \eta_1\om + \eta_2\ka + \cdots | \\
  \approx
  t_0 \left| \frac{\om + \ell_1\ka + t_2\ka^2}{\om + \ell_1\ka + \ell_2\ka^2} \right|
  (1 + \Re\eta_1\,\om + \Re\eta_2\,\ka)
  \quad \text{near $\pairzero$.}
\end{multline}
%
The coefficients have the following significance:
\begin{enumerate}
  \item $\ell_1$ is the rate at which the anomaly in $\omega$ moves past the guided-mode frequency $\omega_0$ as a function of $\kappa$. 
  \item $\lim\limits_{\ka,\om\to0} T(\kappa,\omega) = t_0$.  In particular, $T(\kappa,\omega)$ is continuous at the guided mode pair $\pairzero$.
  \item $\frac{\partial}{\partial\omega} T(\kappa_0,\omega)\!\mid_{\omega_0} \,= t_0 \Re(\eta_1)$.  This relates $\Re\eta_1$ to the slope of the non-resonant transmission at~$\kappa=\kappa_0$.
  \item $\frac{\partial}{\partial\kappa} T(\kappa,\omega_0)\!\mid_{\kappa_0} \,=
  t_0 \Re\!\!\left( \frac{t_2-\ell_2}{\ell_1} + \eta_2 \right)$.
  \item The coefficients $t_2$ and $\ell_2$ control the spreading of the peak and dip of the anomaly as $\kappa$ is perturbed from~$\kappa_0$.
\end{enumerate}

\bigskip 
\noindent
\hfill
\raisebox{1.8em}{\includegraphics[width=0.35\textwidth]{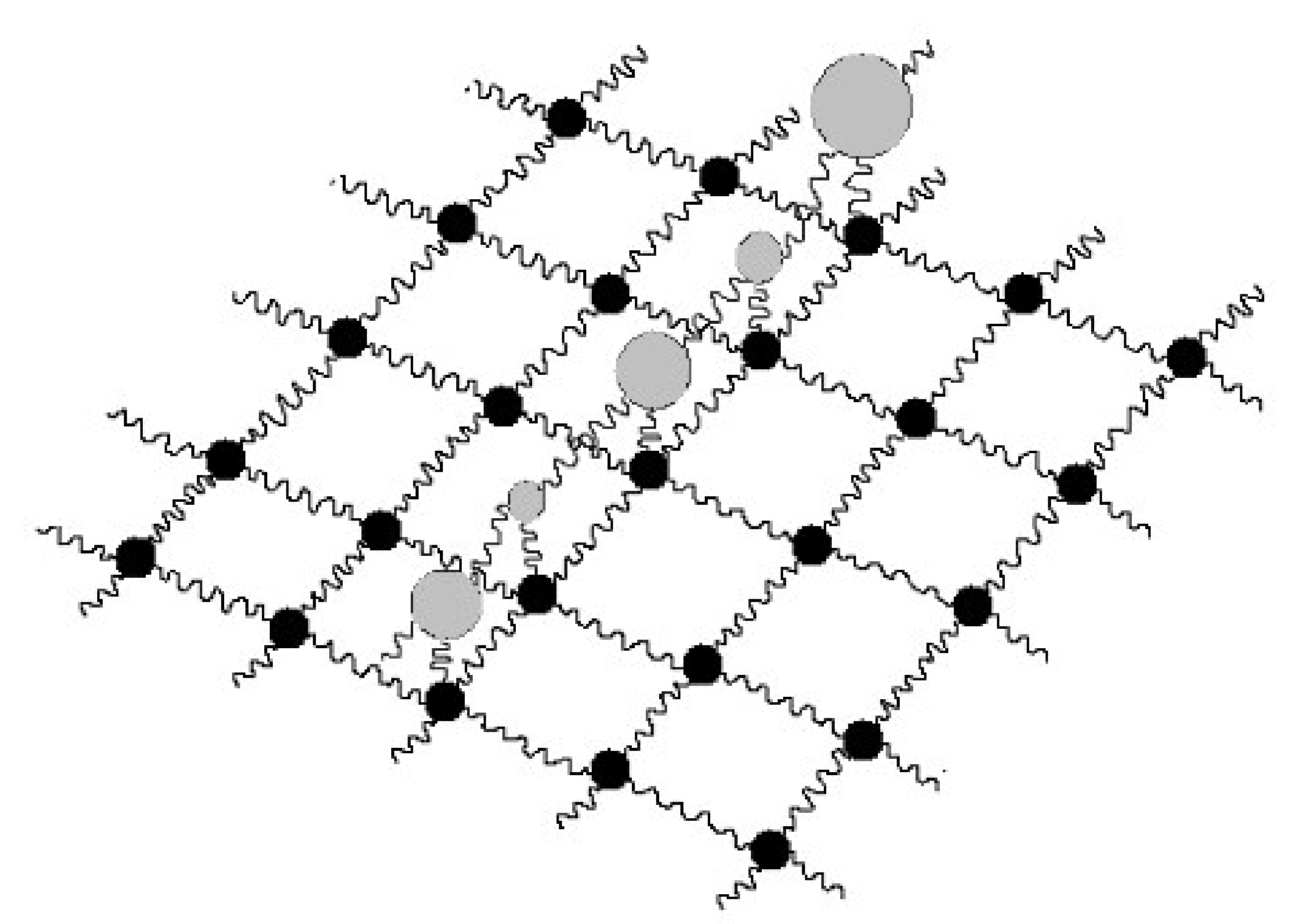}}
\hspace{1ex}
\includegraphics[width=0.49\textwidth]{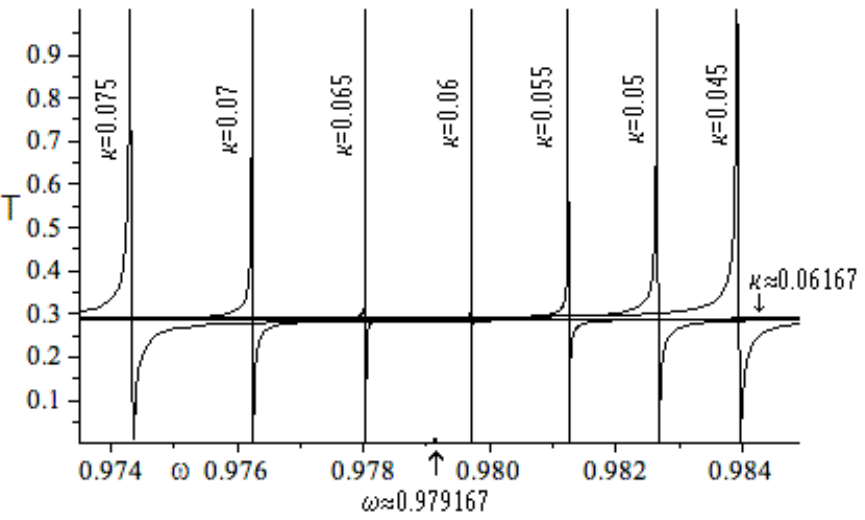}
\hfill\hspace*{0.1ex}

\noindent
 {\small Fig. 1. Plane waves in the uniform 2D lattice are incident upon the coupled periodic 1D lattice from the left and are transmitted to the right.  The square root of the percentage of transmitted energy as a function of frequency $\omega$ is shown for several values of the Bloch wave vector $\kappa$ near the parameters $\pairzero=(0.06167,0.979167)$ of a guided mode.  The exact calculation and the theoretical formula are practically identical.}

\subsection{Case 2: $\ell_1=0$}

This case occurs when the dispersion relation $\ell(\kappa,\omega)=0$, solved for $\omega$, is symmetric in the variable $\kappa$ about $\kappa_0$.  In particular, this occurs at resonant frequencies for $\kappa_0=0$ if the structure has symmetry about a line or plane perpendicular to the plane of the scatterer \cite{Bonnet-BeStarling1994,ShipmanVenakides2005,ShipmanVolkov2007,TikhodeevYablonskiMuljarov2002}, as is the case in Fig. 2.

From the second relation in \eqref{relations1}, we observe that $\ell_1=0$ implies $r_1=0$ and $t_1=0$ also, and we find that the following three terms in the expansion give information about the coefficients:
\begin{equation}\label{relations2}
\renewcommand{\arraystretch}{1.2}
\left.
  \begin{array}{ll}
      \om^2 \;\text{term:} & 1 = r_0^2 + t_0^2, \\
      \om\ka^2 \;\text{term:} & \Re\ell_2 = r_0^2\,\Re r_2 + t_0^2\,\Re t_2, \\
      \ka^4 \;\text{term:} & |\ell_2|^2 = r_0^2\,|r_2|^2 + t_0^2\,|t_2|^2.
  \end{array}
\right.
\end{equation}
Because $\ell_1=0$, the expansions (\ref{form2E},\ref{form2a},\ref{form2b}) reduce to
\begin{eqnarray}
  \ell\pair=0 &\iff& \omega = \omega_0 - \ell_2(\kappa-\kappa_0)^2 - \dots\,, \label{form3E}\\
  a\pair=0 &\iff& \omega = \omega_0 - r_2(\kappa-\kappa_0)^2 - \dots\,, \label{form3a}\\
  b\pair=0 &\iff& \omega = \omega_0 - t_2(\kappa-\kappa_0)^2 - \dots\,. \label{form3b}
\end{eqnarray}
%

The second of the relations \eqref{relations2} tells us that the curve $\ell\pair$ lies between the curves $a\pair\!=\!0$ and $b\pair\!=\!0$.  In particular, if $r_2$ and $t_2$ are real and $\ell_2$ is imaginary, then the transmission peak and dip move in opposite directions away from the guided mode frequency $\omega_0$ as $\kappa$ is perturbed from $\kappa_0$.

In \cite{ShipmanVenakides2005}, we demonstrate that the following approximate expression for the transmission anomaly near $\pairzero$ matches numerical simulations for the scattering of $E$-polarized fields by a periodic lossless dielectric slab:
\begin{equation}\label{FanoPC}
T^2\pair \approx \frac{t_0^2 | \om + t_2\ka^2 |^2 (1+\eta\,\om)^2}
                   {r_0^2 | \om + r_2\ka^2 |^2 + t_0^2 | \om + t_2\ka^2 |^2
                     (1+\eta\,\om)^2} \quad \text{near } (\kappa_0,\omega_0).
\end{equation}
Here, the coefficients have the following significance:
\begin{enumerate}
  \item $\lim\limits_{\omega\to\omega_0} T(\kappa_0,\omega) = t_0$ \;and\; $\displaystyle\lim\limits_{\kappa\to\kappa_0} T(\kappa,\omega_0) = t_0 \left| \frac{t_2}{\ell_2} \right|$.
  \item $\displaystyle \eta = \frac{1}{t_0r_0^2}\frac{\partial T}{\partial\omega}(\kappa_0,\omega)\!\mid_{\omega_0}$.
   \item The coefficients $t_2$ and $r_2$ control the spreading of the peak and dip of the transmission anomaly as $\kappa$ is perturbed from $\kappa_0$.
\end{enumerate}

\bigskip 
\noindent
\centerline{
\raisebox{3ex}{\includegraphics[width=0.28\textwidth]{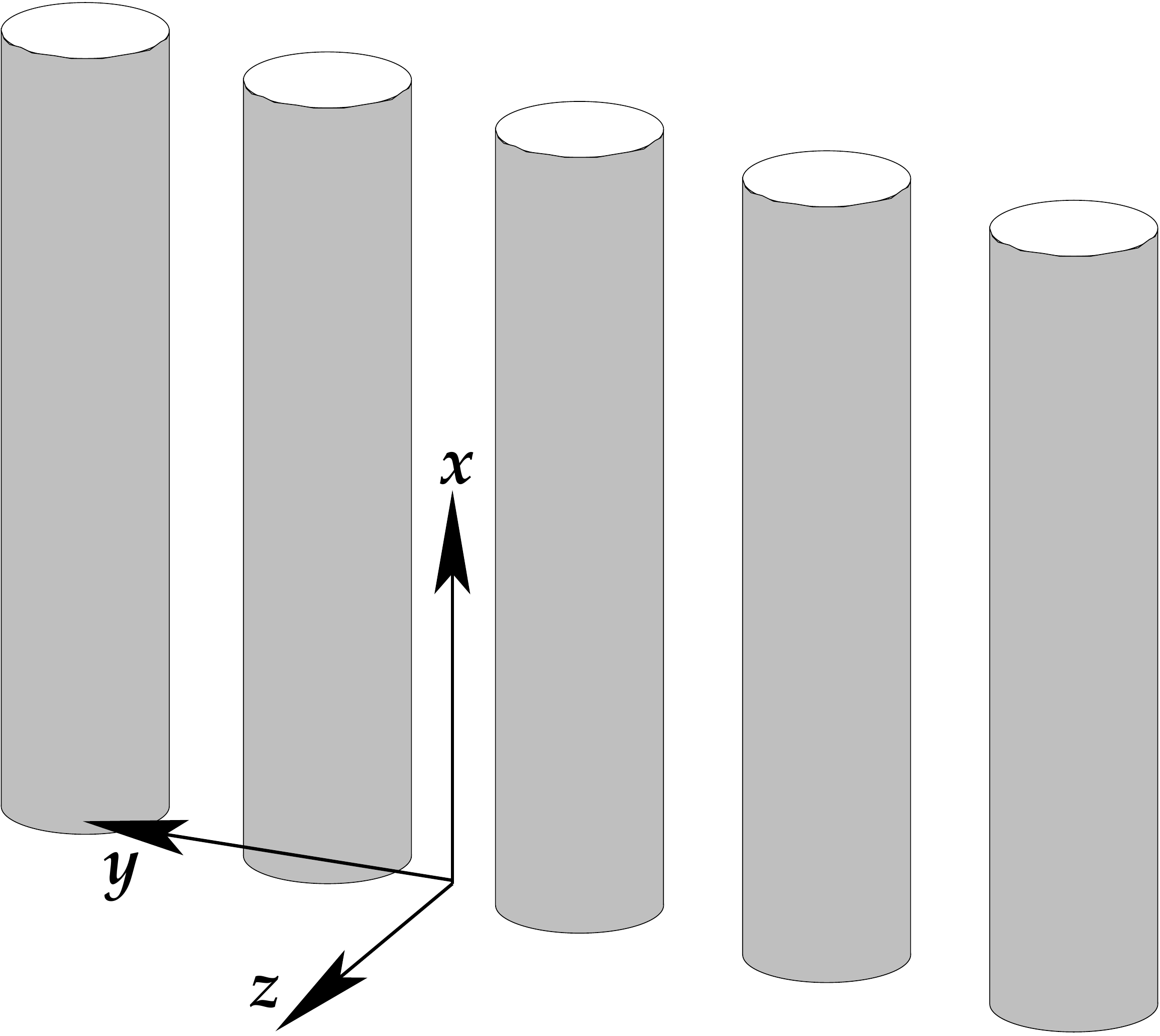}}
\hspace{2em}
\includegraphics[width=0.45\textwidth]{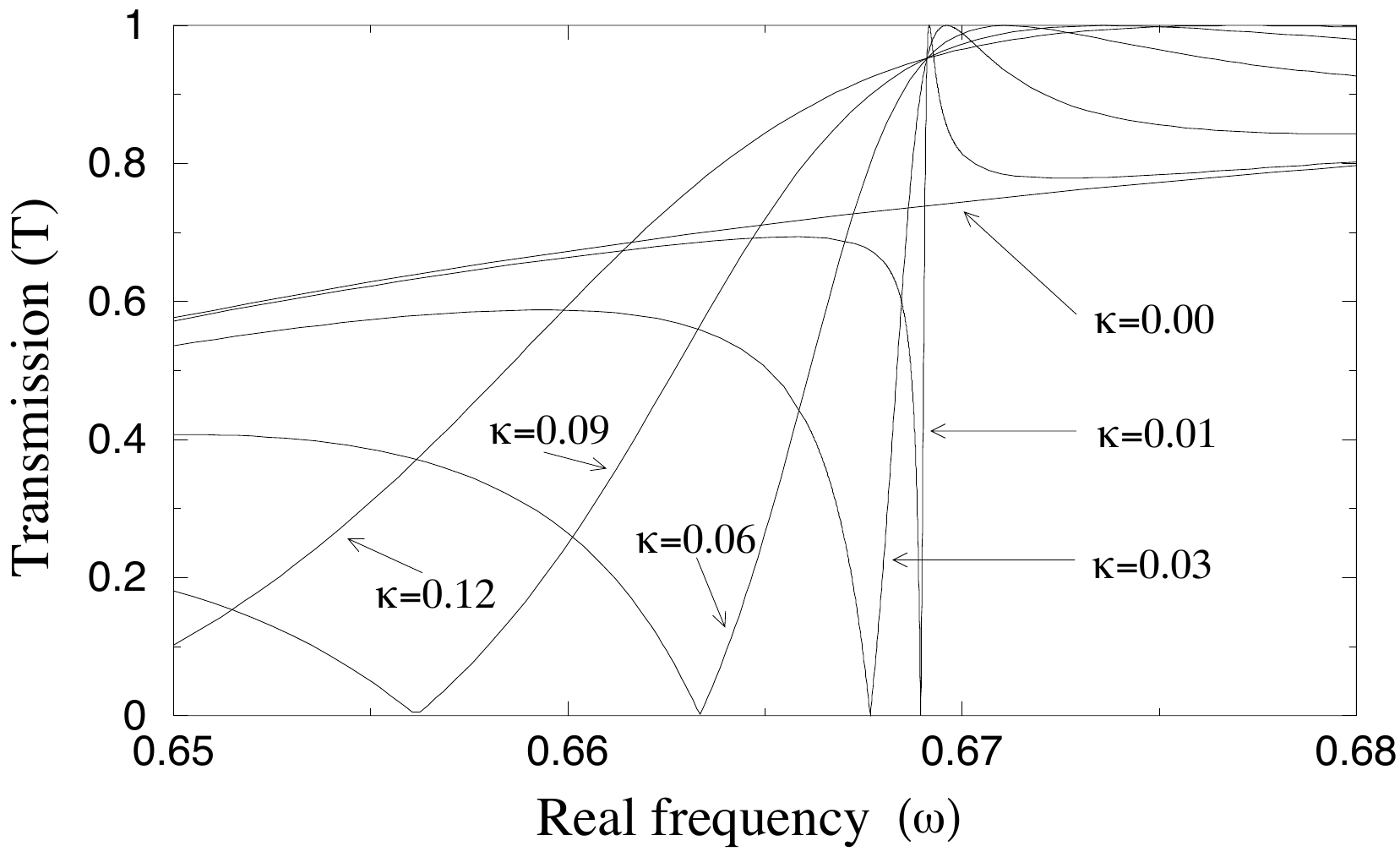}
}

\noindent
{\small Fig. 2. Transmission of E-polarized EM fields through a periodic array of infinitely tall lossless rods ($\epsilon=12$, $\mu=1$) as a function of the reduced frequency $\omega$, for various of values of the wave vector $\kappa$ near the parameters $(0,0.669)$ of a guided mode.  The exact calculation and the theoretical formula are practically identical.
}
\vspace{10pt}

\subsection{Relation to the original Fano shape}

One of the classic examples of anomalous scattering behavior is observed in the excitation of the noble gases near characteristic energies of the atom (the Auger states \cite{ReedSimon1980}).  The anomalies exhibit a peak and a dip and are called ``Fano resonances" after the work of U. Fano \cite{Fano1961}, in which he derived a formula for this shape:
\begin{equation}\label{Fano}
\sigma = \text{const.} \frac{(q + f)^2}{1+f^2}, \quad
       f = \frac{\omega-\omega_\text{res}}{\Gamma/2},
\end{equation}
where the parameters $q$ and $\Gamma$ control the locations of the peak and
dip and $f$ is the deviation of the frequency from resonance normalized to a characteristic width $\Gamma$.

There is a connection between formula \eqref{FanoPC} and the Fano shape \eqref{Fano} that can be expressed in concrete terms.  Namely, the six real parameters in \eqref{FanoPC} ($t_0$, $\eta$, and the real and imaginary parts of $t_2$ and $r_2$) can be reduced to two if certain conditions are satisfied, resulting in the Fano shape.  The conditions are (in addition to $\ell_1=0$)
\begin{enumerate}
   \item  $r_2$ and $t_2$ are real \; (meaning that the extremal values of $T$ are $0$ and $1$),
   \item  $\eta=0$ \; (the background transmission is flat),
   \item  $r_0^2r_2 + t_0^2t_2 = 0$ \; (for small real $\ka$, the dispersion relation given by \eqref{form3E} is purely imaginary).
\end{enumerate}
The resonance \eqref{FanoPC}, as a function of frequency $\omega$, now reduces to the Fano shape \eqref{Fano} with
\begin{equation}
 \Gamma = 2\bar\kappa^2\sqrt{(rr_0)^2 + (tt_0)^2\,} \quad \text{and} \quad q=t/\sqrt{(rr_0)^2 + (tt_0)^2\,}.
\end{equation}
The first condition is satisfied by the example in Fig. 2, while the second and third are not.

\section{Phase anomalies and resonant amplitude enhancement}

The expansions (\ref{form2E},\ref{form2a},\ref{form2b}) can be utilized also to yield a formula for the shape of the argument of the complex transmission coefficient near the guided mode parameters $\pairzero$,
\begin{equation}
  \text{phase of transmitted field} = \arg b\pair - \arg \ell\pair,
\end{equation}
which exhibits sharp spike as a function of $\omega$ for values of $\kappa$ close to $\kappa_0$.  This quantity is closely related to the effective density of states associated with transmission of incident waves through a periodic structure \cite{BendicksoDowlingScalora1996}.

A phenomenon that is directly associated to the interaction of plane waves with guided modes is the resonant amplitude enhancement of the scattered field inside the structure.  The scattering problem is nearly singular for small perturbations, so the solution in the scatterer is large compared to the incident field and resembles a guided mode.  We have established in \cite{ShipmanVenakides2005} a leading-order result for the behavior of this enhancement as a function of~$\kappa$:
\begin{equation}
  \mathrm{amplitude\;enhancement} \sim \frac{\mathrm{const.}}{|\kappa|} \quad (\kappa\to0).
\end{equation}
One of the essential ingredients in the analysis is the fact that there is no resonant enhancement at $\kappa\!=\!\kappa_0$.  This fact is tantamount to the fact that the scattering problem has a solution at the parameters of the guided mode, although it is not unique \cite{Bonnet-BeStarling1994}; in other words, the source field has no ``resonant component" at normal incidence.

\vspace{2ex}
\noindent
{\bf Acknowledgment}

This work was supported by NSF grant DMS-0505833 (S. Shipman and N. Ptitsyna) and NSF grants DMS-0207262 and DMS-0707488 (S. Venakides).

\bibliography{ShipmanArXiv}

\end{document}